\documentclass[aps,prl,twocolumn,showpacs,superscriptaddress]{revtex4-1}

\usepackage{graphicx}
\usepackage{amsmath}
\usepackage{bm}

\newcommand{\mso}{MnSb$_2$O$_6$}
\newcommand{\impurity}{Mn$_2$Sb$_2$O$_7$}
\newcommand{\tn}{\textit{T}$_\mathrm{N}$}

\newcommand{\lsite}{Ba$_3$NbFe$_3$Si$_2$O$_{14}$}

\begin{document}

\title{MnSb$_2$O$_6$: A polar magnet with a chiral crystal structure}

\author{R. D. Johnson}
\email{r.johnson1@physics.ox.ac.uk}
\affiliation{Clarendon Laboratory, Department of Physics, University of Oxford, Oxford, OX1 3PU, United Kingdom}
\affiliation{ISIS facility, Rutherford Appleton Laboratory-STFC, Chilton, Didcot, OX11 0QX, United Kingdom}
\author{K. Cao}
\affiliation{Department of Materials, University of Oxford, Parks Road, Oxford OX1 3PH, United Kingdom}
\author{L. C. Chapon}
\affiliation{Institut Laue-Langevin, BP 156X, 38042 Grenoble, France}
\author{F. Fabrizi}
\author{N. Perks}
\affiliation{Clarendon Laboratory, Department of Physics, University of Oxford, Oxford, OX1 3PU, United Kingdom}
\author{P. Manuel}
\affiliation{ISIS facility, Rutherford Appleton Laboratory-STFC, Chilton, Didcot, OX11 0QX, United Kingdom}
\author{J. J. Yang}
\affiliation{Laboratory for Pohang Emergent Materials, Pohang University of Science and Technology, Pohang 790-784, Korea}
\author{Y. S. Oh}
\author{S-W. Cheong}
\affiliation{Rutgers Center for Emergent Materials and Department of Physics and Astronomy, 136 Frelinghuysen Road, Piscataway 08854, New Jersey, USA.}
\author{P. G. Radaelli}
\affiliation{Clarendon Laboratory, Department of Physics, University of Oxford, Oxford, OX1 3PU, United Kingdom}

\date{\today}

\begin{abstract}
Structural and magnetic chiralities are found to coexist in a small group of materials in which they produce intriguing phenomenologies such as the recently discovered skyrmion phases.  Here, we describe a previously unknown manifestation of this interplay in  \mso, a trigonal oxide with a chiral crystal structure.  Unlike all other known cases, the \mso\ magnetic structure is based on co-rotating cycloids rather than helices.  The coupling to the structural chirality is provided by a magnetic axial vector, related to the so-called vector chirality.  We show that this unique arrangement is the magnetic ground state of the symmetric-exchange Hamiltonian, based on \emph{ab-initio} theoretical calculations of the Heisenberg exchange interactions, and is stabilised by out-of-plane anisotropy. \mso\  is predicted to be multiferroic with a unique ferroelectric switching mechanism.
 
\end{abstract}

\pacs{75.85.+t, 75.25.-j, 75.30.Et}

\maketitle

The search for materials applicable to novel multifunctional solid-state technology has driven the study of exotic electronic ordering phenomena, which arise due to interactions between magnetic, electronic, and structural degrees of freedom. For example, the discovery of skyrmion phases  \cite{muhlbauer09,adams12} has focussed attention on the interplay between complex magnetism and crystal symmetries; in particular, the coupling of structural and magnetic chiralities.  Two archetypes are known:  in MnSi and related metal silicides and germanides,  the magnetic structure inherits the chirality of the acentric crystal structure through antisymmetric Dzyaloshinskii-Moriya (DM) exchange, forming either a simple helix or a more complex ``3-q'' skyrmion phase in applied magnetic fields \cite{muhlbauer09}.  By contrast, the iron langasite (\lsite) \cite{marty08,stock11} magnetic structure is helical, but also possesses a net triangular chirality \cite{tc},  induced by geometrical frustration inherent to the crystal structure.  The magneto-structural coupling, primarily due to symmetric (Heisenberg) exchange, involves structural chirality, magnetic helicity and triangular chirality, which can be combined into a phenomenological invariant.

In this letter we present a third, previously unrecognised example, \mso, which bears similarities and intriguing differences with langasite, and, most strikingly, is predicted to be multiferroic.  Both materials crystallize in the same space group ($P321$), and have similar building blocks and exchange pathways.  However, the \mso\ magnetic structure refined from neutron diffraction data is very different to that of langasite, being based on \emph{cycloids} rather than \emph{helices}.  We show that the two magnetic structures are related by a global spin rotation; both having similar forms of the Heisenberg exchange energy, but stabilised by different anisotropies.   Seven nearest-neighbor (NN) super-super-exchange (SSE) interactions, calculated using \emph{ab-initio} density functional theory (DFT), yield a magnetic ground state and propagation vector in good agreement with experiment. Furthermore, \mso\ is predicted to be weakly polar and a multiferroic of an unusual kind, since reversing the electric field would result in a switch between single- and two-domain configurations.

\begin{figure}
\includegraphics[width=8cm]{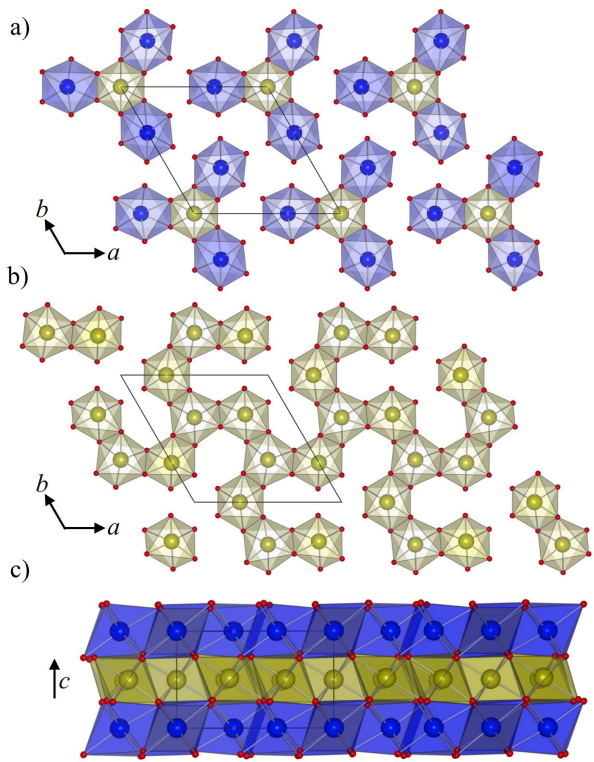}
\caption{\label{crystfig}(Color online) The crystal structure of \mso. (a) $z=0$ layer of manganese triangles with intra-connecting SbO$_6$ octahedra. (b) $z=\tfrac{1}{2}$ layer of SbO$_6$ octahedra. (c) Projection of the crystal looking down the $b^*$ axis. MnO$_6$ and SbO$_6$ octahedra are colored blue and yellow, respectively. Oxygen ions are shown as red spheres and the unit cell is drawn in black.}
\end{figure}

Trigonal \mso\ is structurally related to sodium fluosilicate \cite{vincent87,scott87}. The lattice is populated by a basis of edge-sharing MnO$_6$ and SbO$_6$ distorted octahedra, which form interleaved layers of isolated manganese triangular plaquettes (Figure \ref{crystfig}a, referred to as ``triangles'' hereafter) and depleted honeycomb lattices of antimony ions (Figure \ref{crystfig}b) stacked along the $c$-axis at $z=0$ and $z=\tfrac{1}{2}$, respectively (Figure \ref{crystfig}c).

Manganese is the only magnetic ion in the crystal, and adopts a valence of 2+ giving a high spin, $S=\tfrac{5}{2}$, and orbitally quenched, $L\simeq0$, moment. MnO$_6$ octahedra are isolated and magnetic interactions occur via SSE pathways (Mn-O-O-Mn). Magnetization data showed evidence for short-range magnetic correlations below $\sim200$~K, and long-range antiferromagnetic order below \tn=12.5~K \cite{reimers89}. Neutron powder diffraction (NPD) \cite{reimers89} showed that below \tn\ the manganese magnetic moments had 3D Heisenberg character and rotate according to the incommensurate propagation vector $\boldsymbol{k}=(0.015, 0.015, 0.183)$ in a plane orthogonal to the $a$-axis -- an approximately cycloidal magnetic structure.

In this study, polycrystalline samples of \mso\ were prepared using 99.9\% pure MnCO$_3$ and 99.999\% pure Sb$_2$O$_3$. Stoichiometric amounts  were mixed, ground and pelletized. The pellets were sintered at 1100 $^\circ$C for 10 hours, followed by furnace cooling to room temperature. The process was repeated twice with intermediate grindingÕs. Single crystals of \mso\ with typical dimensions 2.5x1.5x0.5 mm$^3$ were grown by chemical vapor transport using pre-reacted powders and a Cl$_2$ gas agent.

\begin{figure}
\includegraphics[width=8.5cm]{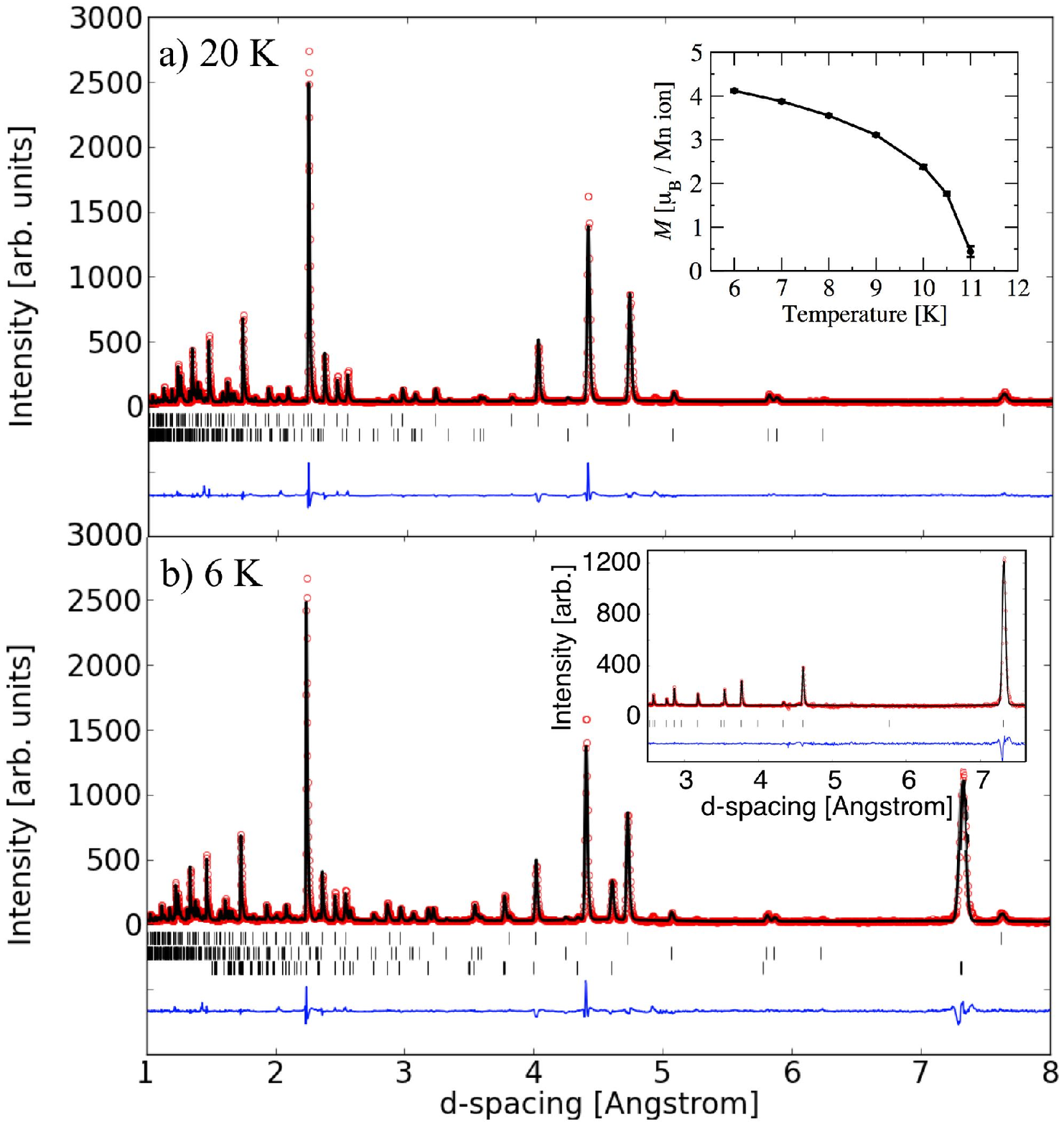}
\caption{\label{npdfig}(Color online) Rietveld refinement against neutron powder diffraction data measured at (a) 20~K, top and bottom tick marks indicate the \mso\ nuclear and impurity \impurity\ peaks, respectively, and at (b) 6~K, top, middle and bottom tick marks indicate the \mso\ nuclear, \impurity, and magnetic peaks, respectively. The measured and calculated profiles are shown with dots and a continuous line, respectively, and a difference curve (observed$-$calculated) is shown at the bottom. Top inset: The temperature dependence of the magnetic moment magnitude. Bottom inset: Data in (a) subtracted from that in (b), showing the magnetic reflections.}
\end{figure}

\begin{figure}
\includegraphics[width=8.5cm]{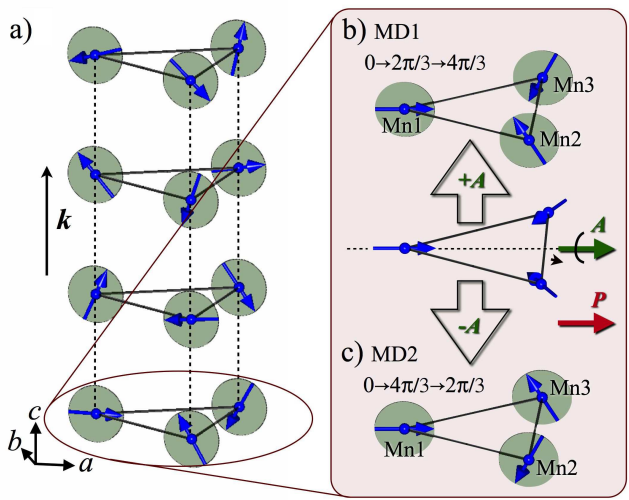}
\caption{\label{magfig}(Color online) The magnetic structure of \mso. For clarity, we only show a single manganese triangle in the $ab$-plane, with the circular spin rotation envelops drawn in grey. a) The cycloid propagating parallel to the $c$-axis. b) and c) The two magnetic domains, MD1 and MD2, respectively, generated by an axial rotation of the totally symmetric co-planar triangular mode.}
\end{figure}

NPD experiments were performed using WISH \cite{chapon11} at ISIS, UK. Comparison of diffraction data (Figure \ref{npdfig}) collected at 20 and 6~K showed a number of reflections evident only below \tn, which could be indexed with propagation vector $\boldsymbol{k}_\mathrm{exp}=(0, 0, 0.1820)$. Contrary to reported results \cite{reimers89}, $\boldsymbol{k}_\mathrm{exp}$ lies along the $\Delta$ line of symmetry, preserving the three-fold rotation as in langasite \cite{marty09}. Apart from this, the published structure \cite{reimers89} provided a good basis for the refinement of our  6~K data using \textsc{fullprof} \cite{rodriguezcarvaja93}, yielding a final reliability factor of $R_{mag}=3.42\%$.  It is not possible to differentiate between amplitude-modulated and rotating magnetic structures by NPD measurements alone, but the former leads to unphysical moment magnitudes and can be excluded.  Manganese moments rotate in a common plane containing $\boldsymbol{k}_\mathrm{exp}$ --- a cycloidal magnetic structure, shown in Figure \ref{magfig}.  The three cycloids in the unit cell have the same polarity, defined as $\boldsymbol{P}_\mathrm{m}=\boldsymbol{k}\times\left(\boldsymbol{S} \times \boldsymbol{S}'\right)$ where $\boldsymbol{S}$ and $\boldsymbol{S}'$ are adjacent Mn spins along the $c$-axis.  This situation should lead to a net ferroelectric polarisation $\perp c$, in analogy to many other cycloidal magnets \cite{kimura03}. The in-plane orientation of the spin rotation plane, and hence the direction of $\boldsymbol{P}_\mathrm{m}$, could not be determined via NPD, but this is not crucial for the interpretation of the underlying physics as will become clear.  Refined moment magnitudes (constrained to be the same for the symmetry-equivalent Mn ions) and relative phases (position within the spin rotation plane) are given in Table \ref{magtab}. Phases between the cycloids on the three symmetry-equivalent Mn atoms are either $0\to \tfrac{2}{3}\pi\to \tfrac{4}{3}\pi$ or $0\to \tfrac{4}{3}\pi\to \tfrac{2}{3}\pi$, as discussed in the Supplementary Material, however, the fit to the powder data is insensitive to this choice of phases, as well as the polarity of the cycloids.

\begin{table}
\caption{\label{magtab}Magnetic structure parameters at 6~K}
\begin{ruledtabular}
\begin{tabular}{c c c c c c c}
Atom &  \textit{x} & \textit{y} & \textit{z} & $|$\textit{\textbf{M}}$|$ ($\mu_\mathrm{B}$) & $\phi_{MD1}$ & $\phi_{MD2}$\\
\hline
Mn1 & 0.6371(9) & 0 & 0 & 4.12(3) & 0 & 0\\
Mn2 & 0 & 0.6371(9) & 0 & 4.12(3) & $\frac{2}{3}$$\pi$ & $\frac{4}{3}$$\pi$\\
Mn3 & -0.6371(9) & -0.6371(9) & 0 & 4.12(3) & $\frac{4}{3}$$\pi$ & $\frac{2}{3}$$\pi$\\
\end{tabular}
\end{ruledtabular}
\end{table}

Unlike the NPD measurement, a single crystal neutron diffraction experiment (performed at D10, ILL and described in the supplementary material), was sensitive to the combination of cycloidal polarity ($\boldsymbol{P}_\mathrm{m}$) and triangular spin arrangement (phases).  To capture all these distinct configurations, we introduce the vector  $\boldsymbol{A}=\boldsymbol{k}\times \boldsymbol{V}$, where $\boldsymbol{V}=\tfrac{1}{3}(\boldsymbol{S}_1 \times \boldsymbol{S}_2 + \boldsymbol{S}_2 \times \boldsymbol{S}_3 +\boldsymbol{S}_3 \times \boldsymbol{S}_1)$ is known as the \emph{vector chirality} and $\boldsymbol{S}_1$, $\boldsymbol{S}_2$, and $\boldsymbol{S}_3$ are spins on the same triangle.   $\boldsymbol{A}$ is an \emph{axial} vector that is necessarily collinear with $\boldsymbol{P}_\mathrm{m}$ and, unlike $\boldsymbol{V}$, is uniquely defined provided that the sense of rotation in traversing the triangle and $\boldsymbol{k}$ are chosen consistently with the right-hand rule. The single crystal unpolarised neutron diffraction intensities depend on the dot product $\boldsymbol{A} \cdot \boldsymbol{P}_\mathrm{m}$, but are insensitive to the individual orientation of $\boldsymbol{A}$ or $\boldsymbol{P}_\mathrm{m}$.  We label the two distinct magnetic configurations with $\boldsymbol{A}$ \emph{parallel} or \emph{antiparallel} to $\boldsymbol{P}_\mathrm{m}$ as MD1 and MD2, respectively (Figure \ref{magfig}b and \ref{magfig}c).

Despite significant differences in magnetic structure, this scenario is strongly reminiscent of langasite, for which the diffraction intensities were sensitive to the product of two \emph{scalar} quantities, $\epsilon_H$ (helicity) and $\epsilon_T$ (scalar triangular chirality) \cite{marty08}, rather than two \emph{vector} quantities as in the present case.   The deep analogy between the two cases becomes apparent when one considers that the \mso\ magnetic structure can be obtained from the \lsite\ magnetic structure by \emph{globally rotating} all spins by 90$^{\circ}$ around $\boldsymbol{A}$, which determines the spin rotation plane, and hence the direction of $\boldsymbol{P}_\mathrm{m}$ (shown in Figure \ref{magfig}). The analogy becomes even more compelling when one considers that both $\boldsymbol{A} \cdot \boldsymbol{P}_\mathrm{m}$ and $\epsilon_H\epsilon_T$ are \emph{pseudo-scalars}, and are therefore capable of providing a coupling to the structural chirality $\sigma$ through the invariants $\sigma \boldsymbol{A} \cdot \boldsymbol{P}_\mathrm{m}$ and $\sigma \epsilon_H\epsilon_T$.  

The best single-domain refinement corresponded to the MD1 magnetic structure, and a significantly improved fit was obtained with a 0.8(1)MD1~:~0.2(1)MD2 domain fraction indicating that, unlike the \lsite\ crystals of the previous studies \cite{marty08,stock11}, our \mso\ crystal was a non-racemic mixture of two chiral structural domains. The inclusion of three-fold symmetry related domains did not improve fitting statistics due to peak intensity averaging inherent to the diffraction experiment.

One would expect the \lsite\ and \mso\ Heisenberg mean-field Hamiltonians to have the same form, since solutions related by a global spin rotation are degenerate in the absence of anisotropy.  However, the number of NN exchange interactions and their magnitudes differ, owing to the different crystal structures. In \lsite\ \cite{marty08,stock11}, the magnetic structure was described by five interactions, J$_1$ -- J$_5$. We adopt the same configuration and labelling scheme, but introduce a further two exchange pathways (J$_6$ and J$_7$) as the larger manganese plaquettes give greater significance to the interactions associated with the J$_2$ triangles.

\begin{figure}
\includegraphics[width=8cm]{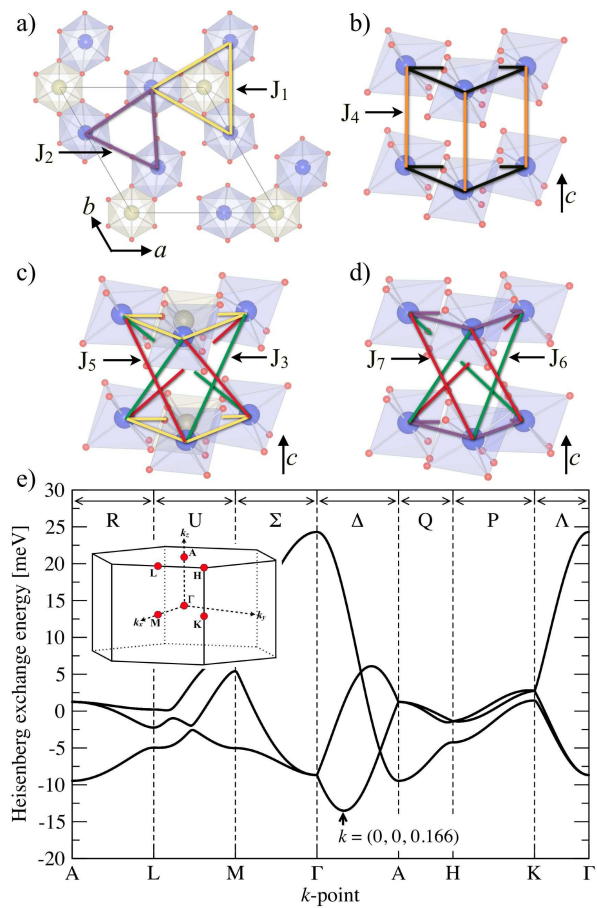}
\caption{\label{pdfig}(Color online) a) J$_1$ (yellow) and J$_2$ (purple) connect triangles of Mn ions centered at the origin, and $(\tfrac{1}{3},\tfrac{2}{3},0)$ and $(\tfrac{2}{3},\tfrac{1}{3},0)$, respectively. b) J$_4$ (orange) connects manganese ions directly above/below one another. c) J$_3$ and J$_5$ are the right-handed (green) and left-handed (red) diagonal pathways of the J$_1$ triangles. Likewise, in d) J$_6$ and J$_7$ are the diagonal pathways of the J$_2$ triangles. e) The Heisenberg exchange energies calculated along the lines of symmetry of the $P321$ Brillouin zone (shown in the inset).}
\end{figure}

The absolute magnitudes of the seven exchange interactions, depicted in Figures \ref{pdfig}a--\ref{pdfig}d, were calculated using DFT within the spin-polarized generalized gradient approximation, implemented in the Vienna \emph{ab-initio} simulations package (VASP) \cite{kresse93,kresse96}. The projector augmented-wave pseudopotentials \cite{blochl94} with a 500 eV plane-wave cutoff were used. A collinear spin approximation without spin-orbit coupling, suitable for calculating symmetric exchanges, gave the values presented in Table \ref{extab}. The paramagnetic region of the magnetic susceptibility measured on the powder sample (not shown here) was fitted to give a Curie-Weiss temperature of $\Theta=-19.6$~K. This value corresponds to a total exchange interaction with an order of magnitude consistent with those calculated.

\begin{table}
\caption{\label{extab}Exchange interactions J$_1$---J$_7$ in units of meV, evaluated through DFT based calculations. Note that the values include the spin component $S(S+1)$.}
\begin{ruledtabular}
\begin{tabular}{c c c c c c c}
J$_1$ & J$_2$ & J$_3$ & J$_4$ & J$_5$ & J$_6$ & J$_7$\\
\hline
0.77 & 1.47 & 2.2 & 1.16 & 0.4 & 1.94 & 0.4\\
\end{tabular}
\end{ruledtabular}
\end{table}

All exchange interactions were found to be significant, however the left-handed interactions, J$_5$ and J$_7$, are weak compared to the right-handed interactions, J$_3$ and J$_6$. It is apparent by qualitative inspection of the magnetic structure that in MD1 J$_3$(J$_6$)~$>>$~J$_5$(J$_7$), and vice-versa for MD2. These left- and right-handed diagonal exchange pathways are directly related to the chirality of the crystal structure, as the inversion symmetry operator that transforms one chiral domain into the other, also interchanges J$_3$(J$_6$) with J$_5$(J$_7$). This demonstrates that the diagonal exchange interactions couple the magnetic domains to the structural chirality.

A mean-field magnetic ground state calculation based upon symmetric Heisenberg exchange was performed for the three manganese sites in the unit cell. The exchange interactions were fixed to the values calculated above, and the propagation vector, $\boldsymbol{k}_\mathrm{calc}$=($k_x, k_y, k_z$), was left free to vary, bounded by the first Brillouin zone of the $P321$ space group. The energy eigenvalues of the Hamiltonian were solved through diagonalization of the Fourier transform of the interaction matrix. A unique minimum energy solution was numerically found corresponding to $\boldsymbol{k}_\mathrm{calc}$=($0, 0, 0.166$). This $\boldsymbol{k}$-vector is in good agreement with that determined experimentally, validating our symmetric exchange model. Importantly, the calculation shows that the magnetic ground state orders with a vector constrained to the $\Delta$ line of symmetry, also consistent with langasite. Figure \ref{pdfig}e shows the Heisenberg exchange energies calculated along all lines of symmetry with the minimum at $\boldsymbol{k}_\mathrm{calc}$ labelled.

The eigenvectors of the Hamiltonian give the phase relationships between the three manganese atoms. By fixing $\boldsymbol{k}_\mathrm{calc}$ we find that the eigenvectors of the minimum energy solution are those corresponding to MD1. Swapping J$_3$(J$_6$) with J$_5$(J$_7$) gives eigenvectors corresponding to MD2, confirming the description given above.

One question remains regarding the anisotropy that favors the cycloidal magnetic structure over a helix in \mso, while the opposite is true of \lsite. Typically rotating magnetic structures are stabilized in acentric crystals (for example MnSi \cite{kataoka81}) via the DM interaction, where energy may be gained through canting spins according to a vector $\boldsymbol{D}$, where $\Delta E_\mathrm{DM}=\sum_n\boldsymbol{D}\cdot(S_n\times S_{n+1})$. However, when $\boldsymbol{D}$ lies perpendicular to a three-fold rotation axis, all vectors will exactly cancel giving no gain in energy. It is likely, therefore, that in \mso\ single-ion-anisotropy favors an out-of-plane spin rotation, breaking the three-fold symmetry. 

To conclude the discussion on the magnetic structure, we consider the possible directions of $\boldsymbol{P}_\mathrm{m}$ and the resulting domain structure.   Although this could not be probed by our diffraction experiment, we expect a unique set of symmetry-equivalent directions for $\boldsymbol{P}_\mathrm{m}$ to be stabilised by magneto-elastic interactions.  The simplest scenario is for $\boldsymbol{P}_\mathrm{m}$ to be \emph{parallel} to one of the 2-fold axes, yielding 3 symmetry-equivalent domains for each structural enantiomer.  It is less likely that $\boldsymbol{P}_\mathrm{m}$ lies in a general in-plane direction, yielding 6 domains (illustrated in the Supplementary Material).  It is important to stress that, in either situation, $\boldsymbol{P}_\mathrm{m}$ and $-\boldsymbol{P}_\mathrm{m}$ are \emph{not} equivalent --- a fact that has a strong bearing on the predicted ferroelectricity, as follows.

We can make a strong prediction that a single structural enantiomer of \mso\ should be ferroelectric, due to the presence of a coupling term $\lambda \boldsymbol{P}_\mathrm{m} \cdot \boldsymbol{P}$ (or $\lambda \sigma \boldsymbol{A} \cdot \boldsymbol{P}$) in the phenomenological free energy expression.   The electrical polarization $\boldsymbol P$ is either parallel or antiparallel to $\boldsymbol{P}_\mathrm{m}$, depending on the sign of the coupling constant $\lambda$. Indeed, by implementing the Berry Phase method \cite{kingsmith93} in our DFT calculations \cite{poldft}, we find an electric polarization of 2 $\mu$Cm$^{-2}$, originating in the DM interaction, and oriented perpendicular to the $c$-axis and within the spin rotation plane. Furthermore, the calculation showed that by switching $\boldsymbol{A}$ the direction of $\boldsymbol{P}$ is reversed. Unfortunately, as the single crystal samples grow as platelets, it is impossible to measure such a small, in-plane ferroelectric polarization reliably. The same is true for powder averaged, polycrystalline measurements. We note that the above coupling is the exact converse to that of ferroaxial multiferroics, in which a \emph{magnetic} chirality couples to a \emph{structural} axial vector to give $\boldsymbol{P}\propto\sigma_m\boldsymbol{A}_s$ \cite{johnson11,johnson12}.  Since $\boldsymbol{P}_m$ and $-\boldsymbol{P}_m$ are \emph{not} equivalent, \mso\ would have a unique ferroelectric switching mechanism, in which reversing the electric field would switch between a single domain and a mixture of at least two domains, discussed further in the Supplementary Material.

\begin{acknowledgments}
The work done at the University of Oxford was funded by an EPSRC grant, number EP/J003557/1, entitled ``New Concepts in Multiferroics and Magnetoelectrics". Work at Rutgers was supported by DOE under Contract No. DE-FG02-07ER46382, and work at Postech was supported by the Max Planck POSTECH/KOREA Research Initiative Program [\#2011-0031558] through the NRF of Korea funded by the Ministry of Education, Science and Technology.
\end{acknowledgments}

\bibliography{mso}

\begin{thebibliography}{20}
\bibitem [{tc()}]{tc}%
  \BibitemOpen
  \href@noop {} {}\bibinfo {note} {Triangular chirality of spins on a triangle,
  $\epsilon_T$, is defined as positive if the spins rotate counterclockwise as
  one circumscribes counterclockwise the triangle with $\boldsymbol{k}$
  pointing towards the observer. Formally,
  $\epsilon_T=\tfrac{1}{3}(\boldsymbol{S}_1\times\boldsymbol{S}_2+\boldsymbol{S}_2\times\boldsymbol{S}_3+\boldsymbol{S}_3\times\boldsymbol{S}_1)\cdot\boldsymbol{k}$.}\BibitemShut
  {Stop}%
\bibitem [{pol()}]{poldft}%
  \BibitemOpen
  \href@noop {} {}\bibinfo {note} {The DFT setup for polarization calculations
  was the same as that for the exchange interactions, except with U=4eV. A
  1$\times$1$\times$3 supercell with $\boldsymbol{k_z}=\tfrac{1}{3}$ was used,
  and the 120$^\circ$ spin structures were set to rotate in the $ac$-plane.
  Spin-orbit coupling was found to induce $\boldsymbol{P} \parallel
  a$.}\BibitemShut {Stop}%
\end{thebibliography}

\end{document}